\documentclass[aps,prl,reprint,groupedaddress]{revtex4-1}

\usepackage{graphicx}
\usepackage{amsmath}
\usepackage{color}
\usepackage{verbatim}
\usepackage{ulem}

\begin{document}


\title{
A Finite-Size Supercell Correction Scheme for Charged Defects in One-Dimensional Systems
}


\author{Sunghyun Kim}
\affiliation{Department of Physics, KAIST, Daejeon 305-701, Korea}

\author{Ji-Sang Park}
\affiliation{National Renewable Energy Laboratory, Golden, Colorado 80401, USA}

\author{K. J. Chang}
\email[]{kjchang@kaist.ac.kr}
\affiliation{Department of Physics, KAIST, Daejeon 305-701, Korea}



\date{\today}

\begin{abstract}
We propose a new finite-size correction scheme for the formation energy of charged defects and impurities
in one-dimensional systems within density functional theory.
The energy correction in a supercell geometry is obtained by solving the Poisson equation in a continuum model
which is described by an anistrotropic permittivity tensor, with the defect charge distribution derived from
first-principles calculations.
We implement our scheme to study impurities and dangling bonds in silicon nanowires and demonstrate 
that the formation energy of charged defects rapidly converges with the supercell size.
\end{abstract}
\maketitle
\section{}
Intrinsic defects commonly exist in materials and extrinsic dopants are indispensable in
applications to devices with specific electronic and optical properties.
Both intrinsic and extrinsic defects can be charged under numerous environments,
such as voltage, temperature, and Fermi level.
First-principles calculations within density functional theory (DFT) have been successful in
describing and predicting the properties of defects \cite{Zhang1991,VandeWalle2004}.
In the DFT approach, the formation energy of a defect mostly relies on a supercell geometry
subject to periodic boundary conditions.
Since a periodic system is used to model the defect, the calculated energy is only meaningful
in the limit where the defect is well isolated.

In the case of charged defects, their formation energies converge much slowly with respect to
the supercell size because of the long-range Coulomb interaction between the defect and its image charges.
Several correction schemes have been proposed for the formation energy of charged defects.
One simple way is to calculate the Madelung-type correction for an array of charged defects with monopole
and quadrupole moments in a neutralizing background \cite{Leslie1985,Makov1995}.
Since this scheme uses a single macroscopic dielectric constant for screening, the convergence problem
still remains for defects in inhomogenous materials.
Recently, Freysoldt \textit{et al.} \cite{Freysoldt2009,Freysoldt2011} have proposed an improved
correction scheme which accounts for the dielectric screening of charged defects in bulk materials
by using the electrostatic potential within DFT and the macroscopic dielectric constant.
Such an approach has been successfully extended to slab systems, which have charged defects 
inside or at surface \cite{Komsa2013}. 

In the potential-based formalism, difficulties still arise in removing the superious electrostatic
interaction of charged defects in complex systems, such as one-dimensional systems embedded in vacuum.
A few theoretical attempts have been made to correct the defect formation energy in nanowires.
The Madelung-type correction was calculated for impurities in Si nanowires by using the dielectric tensor
rather than the dielectric constant \cite{Rurali2009,Rurali2010}.
Since the dielectric tensor cannot properly describe the shape and volume ratio of the embedded media,
this scheme is more appropriate for an anisotropic bulk system.
In other approach \cite{Chan2011}, where the energy is directly calculated in real space, with only
the periodic boundary condition imposed along the wire axis, the Madelung-type correction is not required.
However, the energy convergence with the wire length was shown to be slow, because the electrostatic interaction
of periodic image charges still exists along the wire.

In this Letter, we propose a finite-size correction scheme for charged defects in one-dimensional systems,
which corrects the formation energy in the supercell geometry of a media surrounded with vacuum. 
The electrostatic interaction of periodic image charges is calculated in the model system, based on the
potential-based formalism which employs the dielectric permittivity tensor and the defect charge distribution
derived from DFT calculations.
Successful applications of the scheme are demonstrated for impurities and dangling bonds in Si nanowires.

Our calculations are performed using the generalized gradient approximation (GGA) proposed by Perdew, Burke,
and Ernzerhof (PBE) \cite{Perdew1996} for the exchange-correlation potential within the DFT and the projector
augmented wave psuedopotentials \cite{Blochl1994}, as implemented in the VASP code \cite{Kresse1999}. 
The wave functions are expanded in plane waves with an energy cutoff of 400 eV.
We consider defects, such as B and P impurities and a surface Si dangling bond,
in $\langle111\rangle$-oriented Si nanowires (SiNWs) with the diameter of about 1.3 nm.
Unless a dangling bond is created, all the surface atoms are passivated by hydrogen.
In supercell geometries, the axial lengths ($L_{z}$) range from 9.49 to 37.96 {\AA} and the lateral sizes
($L_{x}$,$L_{y}$) vary from 20 to 40 {\AA} by increasing the vacuum pad.
The $k$-points in the Brillouin zone are generated by the Monkhorst-Pack mesh \cite{Monkhorst1976},
1$\times$1$\times$8 for $L_{z}$ = 9.49 {\AA} and 1$\times$1$\times$2 for $L_{z}$ = 37.96 {\AA}.
All the ionic coordinates are relaxed until the residual forces are less than 0.04 eV/{\AA}.

The periodic images of a defect with charge $q$ cause a spurious interaction. 
The formation energy of the defect can be corrected by the difference in electrostatic energy between
periodic and open boundary conditions \cite{Freysoldt2009}:
\begin{equation}
E_\text{corr}=E_\text{isolated}-E_\text{periodic} +q\Delta V,
\end{equation}
where $q\Delta V$ is added to align the electron chemical potential \cite{Taylor2011}.
In a medium with an anisotropic dielectric permittivity tensor $\boldsymbol{\varepsilon}$, the Poisson equation
is given by
\begin{equation}\label{eq:poisson}
\nabla \cdot \lbrack \boldsymbol{\varepsilon} (\mathbf{r})  \nabla \phi (\mathbf{r}) \rbrack = -\rho (\mathbf{r}),
\end{equation}
where $\rho$ and $\phi$ are the model defect charge distribution and the corresponding model electrostatic potential,
respectively.
In a periodic system, Eq. (\ref{eq:poisson}) can be efficiently solved in momentum space.
Here we focus on a nanowire system with the homogeneous dielectric permittivity along the \textit{z}-direction. 
Thus, $\boldsymbol{\varepsilon}$ depends only on $x$ and $y$ such as 
\begin{equation}
\label{eq:epsilon}
\boldsymbol{\varepsilon}(\mathbf{r}) = 
\left( \begin{array}{ccc}
\varepsilon_{\perp}(x,y) & 0                                         & 0 \\
0                                         & \varepsilon_{\perp}(x,y) & 0 \\
0                                         & 0                                          & \varepsilon_{\parallel}(x,y)
\end{array} \right), 
\end{equation}
where $\varepsilon_{\perp}$ and $\varepsilon_{\parallel}$ are the dielectric permittivities perpendicular
and parallel to the wire axis, respectively.
With Eq. (\ref{eq:epsilon}), the Fourier transform of Eq. (\ref{eq:poisson}) is
\begin{equation}\label{eq:poissonRec}
\begin{split}
\rho (\mathbf{G})& = \sum_{\mathbf{G}^{\prime}} \sum_{i=x,y,z}
G_iG_i^{\prime}\varepsilon_{ii}(\mathbf{G-G^{\prime}})\phi(\mathbf{G}^{\prime})\\
&= \sum_{G_{x}^{\prime},G_{y}^{\prime}}
\phi(G_{x}^{\prime},G_{y}^{\prime},G_{z})
\Bigl[~
G_{z}^{2} 
~\varepsilon_{\parallel} (G_{x}-G_{x}^{\prime}, G_{y}-G_{y}^{\prime}) \\
&
+ 
\left( G_{x}G_{x}^{\prime}+G_{y}G_{y}^{\prime} \right) 
\varepsilon_{\perp}(G_{x}-G_{x}^{\prime}, G_{y}-G_{y}^{\prime})
~\Bigl],
\end{split}
\end{equation}
where $\mathbf{G}$ represents the reciprocal lattice vector in a supercell and $\varepsilon_{ii}$
satisfies the relation, $\varepsilon_{ii}(\mathbf{G}) = \varepsilon_{ii}(G_x,G_y)\delta(G_z)$.
For a given defect charge distribution, we obtain the potential $\phi(\mathbf{G})$ by solving a set of linear equations
in Eq. (\ref{eq:poissonRec}) and finally the electrostatic energy of the periodic system,
\begin{equation}
E_{periodic} = \frac{1}{2} \sum_{\mathbf{G}}\phi(\mathbf{G}) \rho(\mathbf{G}).
\end{equation}

In previous studies \cite{Freysoldt2009,Freysoldt2011,Komsa2013}, an analytic formula or an image charge method
was used to calculate the electrostatic energy $E_\text{isolated}$ of an isolated defect.
However, it is difficult to use such methods for a complex shaped medium.
Here we use a finite volume method (FVM) \cite{Guyer2009} to directly solve the Poisson equation in real space.  
In FVM, Eq. (\ref{eq:poisson}) is discretized by integrating over a subvolume $v$:
\begin{equation}
\oint_s 
\boldsymbol{\varepsilon}(\mathbf{r}) \nabla \phi (\mathbf{r})
\cdot\mathbf{n}\ ds = - \int_v \rho (\mathbf{r})\ dv,
\end{equation}
where $s$ represents the boundary of the subvolume $v$ and $\mathbf{n}$ is the corresponding outward normal vector.
If the Dirichlet boundary condition is adopted, the potential $\phi(\mathbf{r})$ becomes zero at the boundary of a large
simulation cell with the lateral size of a few $\mu$m.
Then, the electrostatic energy is
\begin{equation}
E_\text{isolated} = \frac{1}{2} \sum_{v}  \phi_{v} q_{v},
\end{equation} 
where $\phi_{v}$ is the potential at the center of the subvolume $v$ with the charge $q_{v}$.

\begin{figure}
\includegraphics[width=\columnwidth]{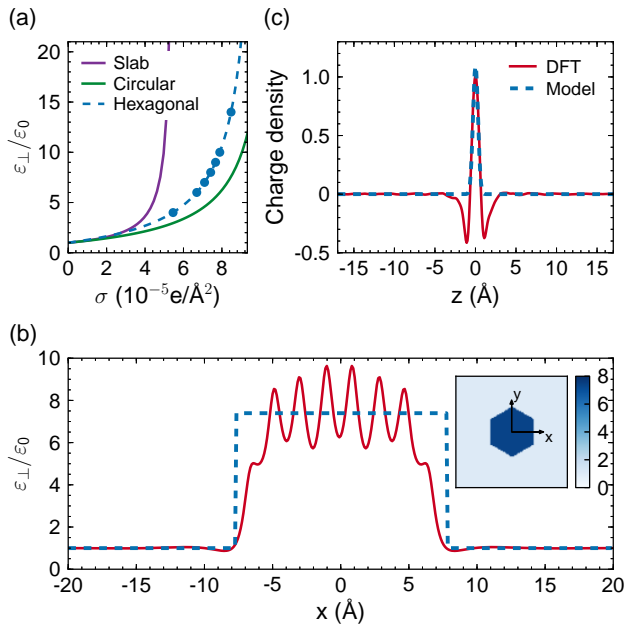}
\caption{\label{fig:epsilon_rho} 
(color online). (a) The results of induced surface charge densities (circles) for model hexagonal nanowires with various dielectric
permittivities $\varepsilon_{\perp}$ under the electric field along the $x$-axis and their best fit to Eq. (\ref{eq:sigma_epsilon})
(dashed line).
(b) The variation of $\varepsilon_{\perp}$ along the $x$-axis based on DFT calculations (red solid line) for a hexagonal
SiNW, the averaged dielectric permittivity (blue dashed line), and its mapping onto the hexagonal cross section (inset). 
(c) The screened defect charge distribution of B$_\text{Si}^{-}$ along the axial direction, which is obtained by integrating
over the cross section (red solid line), and its Gaussian fit for the model charge distribution (blue dashed line). }
\end{figure}

To calculate the electrostatic energies, we need information on the dielectric permittivity and defect charge distribution.
One can use a self-consistent response theory to compute the dielectric permittivity, which is given by the ratio
of screened and external electric fields.
Although this method works well for bulk and slab systems \cite{Freysoldt2011,Komsa2013}, 
it cannot be straightforwardly used for nanowire systems, because the screened electric field in medium
does not scale with $1/{\varepsilon}$.
For nanowires with circular and elliptical cross sections, analytic solutions exist for the electric displacement \cite{Hamel2008}.
However, it is nontrivial to obtain an analytic form if the wire cross section is arbitrary.
Here we consider model nanowires with hexagonal cross sections and numerically derive the relation between the dielectric
permittivity $\varepsilon_{\perp}$ and the induced surface charge density $\sigma$.
We apply an external electric field ${\cal E}_{ext}$ along the $x$-axis, perpendicular to the wire, and calculate
the induced surface charge densities for various dielectric permittivities, as shown in Fig.~\ref{fig:epsilon_rho}(a).
As expected, more charges are induced on the wire surface with increasing of $\varepsilon_{\perp}$.
Because the induced surface charge lies in between those for slab and circular nanowire systems, the results for
$\varepsilon_{\perp}$ and $\sigma$ can be well fitted to the parametrized formula with $a=1.60$ and $b=1.84$, 
\begin{equation}\label{eq:sigma_epsilon} 
\frac {\varepsilon_{\perp}} {\varepsilon_{0}} = \frac {1 + a \sigma / (b\varepsilon_{0} {\cal E}_{ext})} 
{1 - \sigma / {(b\varepsilon_{0} {\cal E}_{ext})}},
\end{equation} 
where $\varepsilon_{0}$ is the dielectric constant of vacuum.
Note that Eq. (\ref{eq:sigma_epsilon}) describes a slab for $a=0$ and $b=1$, whereas it characterize a circular wire
for $a=1$ and $b=2\cos\theta$.
For arbitrary shaped wires, the parameters $a$ and $b$ can be determined by performing similar calculations for
given dielectric permittivities.

The variation of $\varepsilon_{\perp}$ across the wire is obtained through DFT calculations for the
hexagonal SiNW under a finite electric field, which is produced by introducing a dipole in vacuum. 
We calculate the cumulative induced charge densities across the wire and then obtain the dielectric permitivity
$\varepsilon_{\perp}$ from Eq. (\ref{eq:sigma_epsilon}), as shown in Fig.~\ref{fig:epsilon_rho}(b). 
As the induced charges distribute over a few atomic layers near the surface, the dielectric permittivity
grows rapidly from the vacuum region. 
Inside the nanowire, the dielectric permittivity oscillates due to the periodic arrangement of atoms.
To remove such an oscillation, we take the averaged dielectric permittivity inside the NW 
and then map it onto the hexagonal cross section.

On the other hand, the dielectric permittivity $\epsilon_{\parallel}$ parallel to the wire axis is rather easily derived from
an effective medium theory \cite{Rurali2010,Pilania2012}. 
The supercell consists of a vacuum with $\varepsilon_{0}$ and a nanowire with the volume faction $\delta_{NW}$
and the dielectric constants, $\varepsilon_{\perp}^{NW}$ and $\varepsilon_{\|}^{NW}$ perpendicular
and parallel to the wire axis, respectively.
For a circular NW, it is known that the dielectric constants of the effective medium are related to those of the NW in the
Maxwell-Garnett approximation,
\begin{equation}\label{eq:MG_perp}
\frac{\varepsilon_{eff,\perp} - \varepsilon_{0}} {\varepsilon_{eff,\perp} + \varepsilon_{0}}
= \delta_{NW}
\frac{\varepsilon_{\perp}^{NW} - \varepsilon_{0}} {\varepsilon_{\perp}^{NW} + \varepsilon_{0}}
\end{equation}
\begin{equation}\label{eq:MG_para}
\varepsilon_{eff,\parallel} 
=  \delta_{NW} (\varepsilon_{\parallel}^{NW} - \varepsilon_{0}) + \varepsilon_{0}.
\end{equation}
In the supercell geometry of SiNW, the effective dielectric constants perpendicular and parallel to the wire axis,
denoted as $\varepsilon_{eff,\perp}$ and $\varepsilon_{eff,\parallel}$, respectively, are directly calculated
by using the Berry-phase formulation of polarization within DFT \cite{Nunes2001,Souza2002}.
Taking the averaged value of $\varepsilon_{\perp}$ inside the NW [Fig.~\ref{fig:epsilon_rho}(b)] as
$\varepsilon_{\perp}^{NW}$, we estimate $\delta_{NW}$ to be 0.126 for $L_{x}$, $L_{y}$= 40 {\AA}
from Eq. (\ref{eq:MG_perp}).
This NW volume is very close to that estimated from the charge density profile, justifying the effective medium theory.
Using the DFT values for $\delta_{NW}$ and $\varepsilon_{eff,\parallel}$, $\varepsilon_{\parallel}^{NW}$
is obtained from Eq. (\ref{eq:MG_para}) and the dielectric permittivity $\varepsilon_{\parallel}$ in Eq. (\ref{eq:epsilon}) 
is finally derived by assigning $\varepsilon_{\parallel}^{NW}$ and $\varepsilon_{0}$ to the NW and
vacuum, respectively.

A Gaussian charge has been used to model the defect charge distribution
in bulk and slab systems \cite{Freysoldt2009,Freysoldt2011,Komsa2012,Komsa2013}. 
The model Gaussian charge, which is obtained by fitting the defect wavefunction, works well for localized defects,
despite the oscillating behavior of the screened charge distribution in local atomic details.
However, the wavefunction approach leads to large errors for delocalized defects, while the defect formation energy
improves by including an exponential tail in the model charge \cite{Freysoldt2011,Komsa2012}.
Since the wavefunction approach only accounts for electrons, its charge distribution is not suitable for
impurities, such as B$_\text{Si}^{-}$ and P$_\text{Si}^{+}$, which have different ionic charges,
as compared to Si. 
Here we derive the model charge distribution from a Gaussian fitting to the total screened charge distribution,
which is the solution of Eq. (\ref{eq:poisson}) for the DFT difference potential,
with $\boldsymbol{\varepsilon}(\mathbf{r}) = \varepsilon_0$.
The DFT difference potential is obtained from the difference in the screened local potential between two supercells
with and without a charged defect.
Since the DFT difference potential consists of the ionic psuedopotential, Hartree potential, and
exchange-correlation potential, the total screened DFT charge distribution $\rho_\text{screened}^\text{(total)}$
is decomposed as 
\begin{equation}
\rho_\text{screened}^\text{(total)} = \rho_\text{bare}^\text{(defect)} + \rho_\text{scr}^\text{(loc)}
+ \rho_\text{scr}^\text{(deloc)},
\end{equation}
where $\rho_\text{bare}^\text{(defect)}$, $\rho_\text{scr}^\text{(loc)}$, and $\rho_\text{scr}^\text{(deloc)}$
are the bare defect charge, localized screening charge, and delocalized screening charge densities, respectively.
For B$_\text{Si}^{-}$ and P$_\text{Si}^{+}$ impurities, $\rho_\text{bare}^\text{(defect)}$ represents the bare
charges of the impurity ions.
In a wire structure, $\rho_\text{scr}^\text{(deloc)}$ corresponds to the induced surface charge.
Fig.~\ref{fig:epsilon_rho}(c) shows the distribution of $\rho_\text{screened}^\text{(total)}$ along the axis
for a B$_\text{Si}^{-}$ defect in the SiNW, which is obtained by integrating over the wire cross section.
In this case, with choosing a proper cutoff radius between the SiNW radius and the Thomas-Fermi screening length of
about 2.3 {\AA} in Si \cite{Resta1977}, the contribution of $\rho_\text{scr}^\text{(deloc)}$ by the induced surface charge is excluded.
It is clear that the charge distribution oscillates in the core region due to two terms, $\rho_\text{bare}^\text{(defect)}$ and
$\rho_\text{scr}^\text{(loc)}$.
We point out that only the charge distribution around the major peak is needed to determine the spatial extent
in the Gaussian fitting in Fig.~\ref{fig:epsilon_rho}(c).
The agreement of the DFT difference potential with the model potential, which is derived from the model
Gaussian charge, is a precursor of justifying the corrected formation energy, as will be discussed below.

\begin{figure}
\includegraphics[width=\columnwidth]{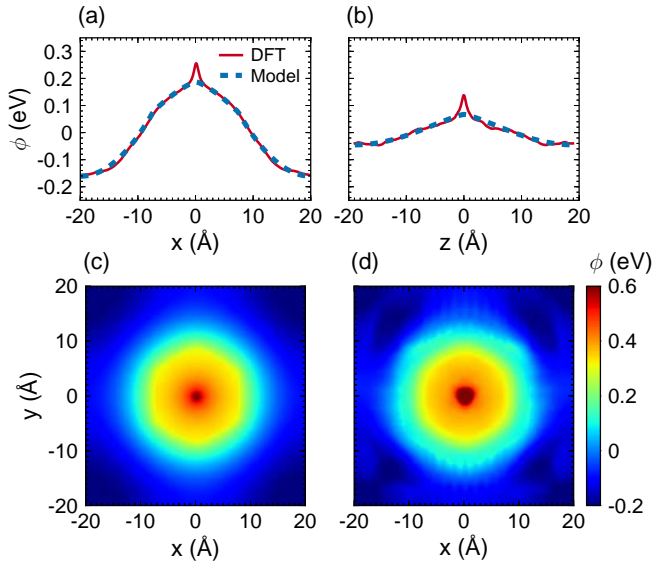}
\caption{\label{fig:potential} 
(color online). Comparison of the model potential (blue dashed line) with the DFT difference potential (red solid line)
for a B$_\text{Si}^-$ impurity positioned at the wire center. In (a)-(b), the potential is averaged over the
$yz$- and $xy$-planes, respectively. Contour plots of (c) the model potential and (d) the DFT difference potential,
which are averaged along the axis. }
\end{figure}

\begin{figure}
\includegraphics[width=\columnwidth]{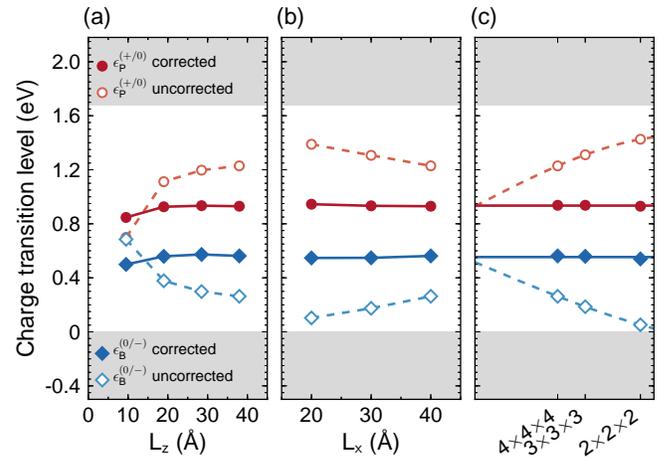}
\caption{\label{fig:CTLconvergence}
(color online). The uncorrected (empty symbols) and corrected (filled symbols) transition levels
of the B (blue diamonds) and P (red circles) impurities are compared for various supercell sizes.
In (a) and (b), the lateral and axial cell sizes are set to be 40 and 37.96 {\AA}, respectively. 
Lines are a guide to the eye. 
In (c), the supercell size is expressed as a multiple of the unit size, 10$\times$10$\times$9.49~{\AA}$^3$.
The extrapolated values (dashed lines) of the uncorrected levels are obtained by fitting to
$a/{\Omega} + b/{\Omega}^{1/3} + c$, where $\Omega$ is the supercell volume,
whereas the average values of the corrected transition levels are drawn by solid lines.
The shaded regions represent the conduction and valance bands of the H-terminated SiNW. }
\end{figure}

Our correction scheme is implemented to calculate the accurate charge transition levels of B and P impurities.
For a charged impurity in the innermost position, we derive the model potential for the model Gaussian charge
from Eq. (\ref{eq:poissonRec}), using the dielectric permittivity tensor.  
We test a B$_\text{Si}^{-}$ impurity in the unrelaxed supercell and find that the model potential agrees well
with the DFT difference potential in Fig.~\ref{fig:potential}, verifying the model Gaussian charge.
We calculate the formation energies of B and P impurities in neutral and charged states for
various supercells, in which the ions are fully relaxed.
We then determine the charge transition levels, $\epsilon_{\text{B}}^{(0/-)}$ for B and
$\epsilon_{\text{P}}^{(+/0)}$ for P, which are defined as the position of the Fermi level where the defect
charge state changes \cite{Zhang1991,VandeWalle2004}.
The corrected and uncorrected charge transition levels are compared in Fig.~\ref{fig:CTLconvergence}.
Without the finite-size corrections, the transition levels do not converge even if the supercell is enlarged
up to 40 {\AA} along the lateral and axial directions. 
On the other hand, the corrected transition levels are nearly flat for either the lateral or axial expansion of the supercell.
In Fig.~\ref{fig:CTLconvergence}(c), the uncorrected transition levels are plotted for the supercells
with the same scaled dimensions along the lateral and axial directions, similar to the previous work \cite{Komsa2013}.
The extrapolated values differ only by 4 and 38 meV from the corrected transition levels for
$\epsilon_{\text{B}}^{(0/-)}$ and $\epsilon_{\text{P}}^{(+/0)}$, respectively,
which are estimated to be 0.55 eV above the valence band maximum and 0.74 eV below the conduction band minimum.
 
For an Al impurity in SiNW, previous studies showed that a vacuum sheath of about 50 {\AA} and
a wire length of about 60 {\AA} were required for satisfactory convergence, with including the
Madelung-type correction in supercell calculations \cite{Rurali2009, Rurali2010}.
In a real-space approach, which does not require the Madelung correction but employs
the periodic boundary condition only along the axis, the converged result was not obtained even with the
wire length of 60 {\AA} \cite{Chan2011}.
In our scheme, the supercell size of about 20 {\AA} is shown to be sufficient to provide converged
transition levels.

\begin{figure}
\includegraphics[width=\columnwidth]{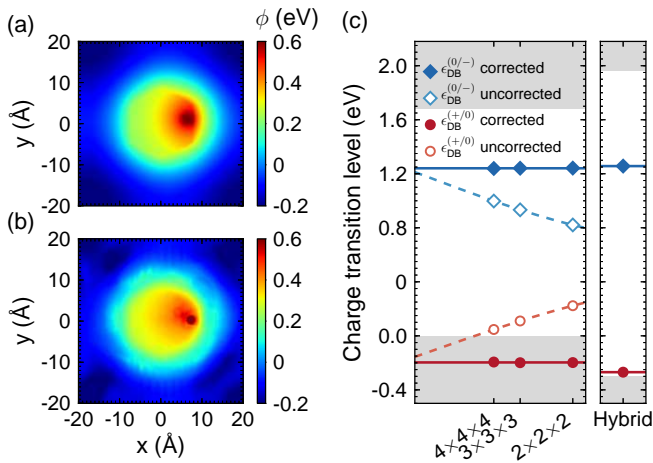}
\caption{\label{fig:DB_potential_CTL}
(color online). Contour plots of (a) the model potential and (b) the DFT difference potential, which are averaged
along the axis, for a surface DB defect. 
(c) The uncorrected (empty symbols) and corrected (filled symbols) transition levels of the DB for various
supercell sizes, with the same sizes and notations as those in Fig.~\ref{fig:CTLconvergence}.
The right panel shows the results of hybrid functional calculations for the supercell of
20$\times$20$\times$18.98 $\text{\AA}^3$. }
\end{figure}

Finally, we consider a localized dangling-bond (DB) defect on the wire surface.
The transition levels of the DB defect in SiNW were previously calculated \cite{Kagimura2007, Hong2010}.
However, the results were erroneous because finite-size corrections were not included.
Especially, the $\epsilon_\text{DB}^{(+/0)}$ level was shown to lie blow the valence band edge,
inferring that DBs hardly capture hole carriers, in contradiction to experiments \cite{Ragnarsson2000}.
Since the DB defect is positioned at the surface site, the effect of the dielectric permittivity tensor becomes
more significant than for defects inside the wire. 
As illustrated in Fig.~\ref{fig:DB_potential_CTL}, the model potential successfully captures the feature
of the DFT difference potential,
indicating that our scheme is applicable for a variety of defects. 
It is clear that both the $\epsilon_\text{DB}^{(+/0)}$ and $\epsilon_\text{DB}^{(0/-)}$
levels converge rapidly, with including the finite-size corrections [Fig.~\ref{fig:DB_potential_CTL}(c)]. 
The corrected $\epsilon_\text{DB}^{(+/0)}$ and $\epsilon_\text{DB}^{(0/-)}$ levels are positioned
at $-0.20$ and 1.24 eV with respect to the valance band maximum, respectively, and these results are similar
to the extraplated values of the uncorrected levels.
Since the band gap is underestimated by the GGA,
the $\epsilon_\text{DB}^{(+/0)}$ level lies within the valence band.
To improve the band gap and transition levels, we additionally perform PBE0 hybrid functional
calculations \cite{Perdew1996a}.
For the supercell of 20$\times$20$\times$18.98 $\text{\AA}^3$, the mixing fraction of the exact
Hartree-Fock exchange is set to be $\alpha$ = 0.11, which is the same as that used for bulk Si \cite{Broqvist2008}.
With the finite-size corrections by GGA, the $\epsilon_\text{DB}^{(+/0)}$ and $\epsilon_\text{DB}^{(0/-)}$
are found to be 0.03 and 1.56 eV above the valence band edge, respectively.
Thus, the hole capture behavior of DBs is more successfully explained.
If an oxide shell is formed in the SiNW, the $\epsilon_\text{DB}^{(+/0)}$ level will be deeper
because the interaction with the image charge is reduced.

In summary, we have developed a posterior correction scheme for calculating the accurate formation
energies of charged defects in one-dimensional systems embedded in vacuum. 
The model potential, which is derived from the Gaussian defect charge, agrees well with
the DFT difference potential, justifying the correction scheme. 
In H-terminated SiNWs, our scheme has shown rapid convergence with respect to the wire length
and the vacuum pad in the formation energies of charged defects, such as B and P impurities
in the innermost position and a localized dangling bond on the surface.
The scheme is applicable to any charged defect in systems with anisotropic dielectric permitivity tensors.  

\begin{acknowledgments}
This work was supported by the National Research Foundation of Korea under Grant No. NRF-2013-077644. 
\end{acknowledgments}

\bibliography{biblio}

\begin{thebibliography}{27}%
\makeatletter
\providecommand \@ifxundefined [1]{%
 \@ifx{#1\undefined}
}%
\providecommand \@ifnum [1]{%
 \ifnum #1\expandafter \@firstoftwo
 \else \expandafter \@secondoftwo
 \fi
}%
\providecommand \@ifx [1]{%
 \ifx #1\expandafter \@firstoftwo
 \else \expandafter \@secondoftwo
 \fi
}%
\providecommand \natexlab [1]{#1}%
\providecommand \enquote  [1]{``#1''}%
\providecommand \bibnamefont  [1]{#1}%
\providecommand \bibfnamefont [1]{#1}%
\providecommand \citenamefont [1]{#1}%
\providecommand \href@noop [0]{\@secondoftwo}%
\providecommand \href [0]{\begingroup \@sanitize@url \@href}%
\providecommand \@href[1]{\@@startlink{#1}\@@href}%
\providecommand \@@href[1]{\endgroup#1\@@endlink}%
\providecommand \@sanitize@url [0]{\catcode `\\12\catcode `\$12\catcode
  `\&12\catcode `\#12\catcode `\^12\catcode `\_12\catcode `\%12\relax}%
\providecommand \@@startlink[1]{}%
\providecommand \@@endlink[0]{}%
\providecommand \url  [0]{\begingroup\@sanitize@url \@url }%
\providecommand \@url [1]{\endgroup\@href {#1}{\urlprefix }}%
\providecommand \urlprefix  [0]{URL }%
\providecommand \Eprint [0]{\href }%
\providecommand \doibase [0]{http://dx.doi.org/}%
\providecommand \selectlanguage [0]{\@gobble}%
\providecommand \bibinfo  [0]{\@secondoftwo}%
\providecommand \bibfield  [0]{\@secondoftwo}%
\providecommand \translation [1]{[#1]}%
\providecommand \BibitemOpen [0]{}%
\providecommand \bibitemStop [0]{}%
\providecommand \bibitemNoStop [0]{.\EOS\space}%
\providecommand \EOS [0]{\spacefactor3000\relax}%
\providecommand \BibitemShut  [1]{\csname bibitem#1\endcsname}%
\let\auto@bib@innerbib\@empty
\bibitem [{\citenamefont {Zhang}\ and\ \citenamefont
  {Northrup}(1991)}]{Zhang1991}%
  \BibitemOpen
  \bibfield  {author} {\bibinfo {author} {\bibfnamefont {S.~B.}\ \bibnamefont
  {Zhang}}\ and\ \bibinfo {author} {\bibfnamefont {J.~E.}\ \bibnamefont
  {Northrup}},\ }\href {\doibase 10.1103/PhysRevLett.67.2339} {\bibfield
  {journal} {\bibinfo  {journal} {Phys. Rev. Lett.}\ }\textbf {\bibinfo
  {volume} {67}},\ \bibinfo {pages} {2339} (\bibinfo {year}
  {1991})}\BibitemShut {NoStop}%
\bibitem [{\citenamefont {{Van de Walle}}\ and\ \citenamefont
  {{Neugebauer}}(2004)}]{VandeWalle2004}%
  \BibitemOpen
  \bibfield  {author} {\bibinfo {author} {\bibfnamefont {C.~G.}\ \bibnamefont
  {{Van de Walle}}}\ and\ \bibinfo {author} {\bibfnamefont {J.}~\bibnamefont
  {{Neugebauer}}},\ }\href {\doibase 10.1063/1.1682673} {\bibfield  {journal}
  {\bibinfo  {journal} {J. Appl. Phys.}\ }\textbf {\bibinfo {volume} {95}},\
  \bibinfo {pages} {3851} (\bibinfo {year} {2004})}\BibitemShut {NoStop}%
\bibitem [{\citenamefont {Leslie}\ and\ \citenamefont
  {Gillan}(1985)}]{Leslie1985}%
  \BibitemOpen
  \bibfield  {author} {\bibinfo {author} {\bibfnamefont {M.}~\bibnamefont
  {Leslie}}\ and\ \bibinfo {author} {\bibfnamefont {N.~J.}\ \bibnamefont
  {Gillan}},\ }\href {\doibase 10.1088/0022-3719/18/5/005} {\bibfield
  {journal} {\bibinfo  {journal} {J. Phys. C}\ }\textbf {\bibinfo {volume}
  {973}},\ \bibinfo {pages} {973} (\bibinfo {year} {1985})}\BibitemShut
  {NoStop}%
\bibitem [{\citenamefont {Makov}\ and\ \citenamefont
  {Payne}(1995)}]{Makov1995}%
  \BibitemOpen
  \bibfield  {author} {\bibinfo {author} {\bibfnamefont {G.}~\bibnamefont
  {Makov}}\ and\ \bibinfo {author} {\bibfnamefont {M.~C.}\ \bibnamefont
  {Payne}},\ }\href {\doibase 10.1103/PhysRevB.51.4014} {\bibfield  {journal}
  {\bibinfo  {journal} {Phys. Rev. B}\ }\textbf {\bibinfo {volume} {51}},\
  \bibinfo {pages} {4014} (\bibinfo {year} {1995})}\BibitemShut {NoStop}%
\bibitem [{\citenamefont {Freysoldt}\ \emph {et~al.}(2009)\citenamefont
  {Freysoldt}, \citenamefont {Neugebauer},\ and\ \citenamefont {{Van de
  Walle}}}]{Freysoldt2009}%
  \BibitemOpen
  \bibfield  {author} {\bibinfo {author} {\bibfnamefont {C.}~\bibnamefont
  {Freysoldt}}, \bibinfo {author} {\bibfnamefont {J.}~\bibnamefont
  {Neugebauer}}, \ and\ \bibinfo {author} {\bibfnamefont {C.~G.}\ \bibnamefont
  {{Van de Walle}}},\ }\href {\doibase 10.1103/PhysRevLett.102.016402}
  {\bibfield  {journal} {\bibinfo  {journal} {Phys. Rev. Lett.}\ }\textbf
  {\bibinfo {volume} {102}},\ \bibinfo {pages} {016402} (\bibinfo {year}
  {2009})}\BibitemShut {NoStop}%
\bibitem [{\citenamefont {Freysoldt}\ \emph {et~al.}(2011)\citenamefont
  {Freysoldt}, \citenamefont {Neugebauer},\ and\ \citenamefont {{Van de
  Walle}}}]{Freysoldt2011}%
  \BibitemOpen
  \bibfield  {author} {\bibinfo {author} {\bibfnamefont {C.}~\bibnamefont
  {Freysoldt}}, \bibinfo {author} {\bibfnamefont {J.}~\bibnamefont
  {Neugebauer}}, \ and\ \bibinfo {author} {\bibfnamefont {C.~G.}\ \bibnamefont
  {{Van de Walle}}},\ }\href {\doibase 10.1002/pssb.201046289} {\bibfield
  {journal} {\bibinfo  {journal} {Phys. Status Solidi B}\ }\textbf {\bibinfo
  {volume} {248}},\ \bibinfo {pages} {1067} (\bibinfo {year}
  {2011})}\BibitemShut {NoStop}%
\bibitem [{\citenamefont {Komsa}\ and\ \citenamefont
  {Pasquarello}(2013)}]{Komsa2013}%
  \BibitemOpen
  \bibfield  {author} {\bibinfo {author} {\bibfnamefont {H.-P.}\ \bibnamefont
  {Komsa}}\ and\ \bibinfo {author} {\bibfnamefont {A.}~\bibnamefont
  {Pasquarello}},\ }\href {\doibase 10.1103/PhysRevLett.110.095505} {\bibfield
  {journal} {\bibinfo  {journal} {Phys. Rev. Lett.}\ }\textbf {\bibinfo
  {volume} {110}},\ \bibinfo {pages} {095505} (\bibinfo {year}
  {2013})}\BibitemShut {NoStop}%
\bibitem [{\citenamefont {Rurali}\ and\ \citenamefont
  {Cartoix\`{a}}(2009)}]{Rurali2009}%
  \BibitemOpen
  \bibfield  {author} {\bibinfo {author} {\bibfnamefont {R.}~\bibnamefont
  {Rurali}}\ and\ \bibinfo {author} {\bibfnamefont {X.}~\bibnamefont
  {Cartoix\`{a}}},\ }\href {\doibase 10.1021/nl802847p} {\bibfield  {journal}
  {\bibinfo  {journal} {Nano Lett.}\ }\textbf {\bibinfo {volume} {9}},\
  \bibinfo {pages} {975} (\bibinfo {year} {2009})}\BibitemShut {NoStop}%
\bibitem [{\citenamefont {Rurali}\ \emph {et~al.}(2010)\citenamefont {Rurali},
  \citenamefont {Palummo},\ and\ \citenamefont {Cartoix\`{a}}}]{Rurali2010}%
  \BibitemOpen
  \bibfield  {author} {\bibinfo {author} {\bibfnamefont {R.}~\bibnamefont
  {Rurali}}, \bibinfo {author} {\bibfnamefont {M.}~\bibnamefont {Palummo}}, \
  and\ \bibinfo {author} {\bibfnamefont {X.}~\bibnamefont {Cartoix\`{a}}},\
  }\href {\doibase 10.1103/PhysRevB.81.235304} {\bibfield  {journal} {\bibinfo
  {journal} {Phys. Rev. B}\ }\textbf {\bibinfo {volume} {81}},\ \bibinfo
  {pages} {235304} (\bibinfo {year} {2010})}\BibitemShut {NoStop}%
\bibitem [{\citenamefont {Chan}\ \emph {et~al.}(2011)\citenamefont {Chan},
  \citenamefont {Zhang},\ and\ \citenamefont {Chelikowsky}}]{Chan2011}%
  \BibitemOpen
  \bibfield  {author} {\bibinfo {author} {\bibfnamefont {T.-L.}\ \bibnamefont
  {Chan}}, \bibinfo {author} {\bibfnamefont {S.}~\bibnamefont {Zhang}}, \ and\
  \bibinfo {author} {\bibfnamefont {J.}~\bibnamefont {Chelikowsky}},\ }\href
  {\doibase 10.1103/PhysRevB.83.245440} {\bibfield  {journal} {\bibinfo
  {journal} {Phys. Rev. B}\ }\textbf {\bibinfo {volume} {83}},\ \bibinfo
  {pages} {245440} (\bibinfo {year} {2011})}\BibitemShut {NoStop}%
\bibitem [{\citenamefont {Perdew}\ \emph
  {et~al.}(1996{\natexlab{a}})\citenamefont {Perdew}, \citenamefont {Burke},\
  and\ \citenamefont {Ernzerhof}}]{Perdew1996}%
  \BibitemOpen
  \bibfield  {author} {\bibinfo {author} {\bibfnamefont {J.~P.}\ \bibnamefont
  {Perdew}}, \bibinfo {author} {\bibfnamefont {K.}~\bibnamefont {Burke}}, \
  and\ \bibinfo {author} {\bibfnamefont {M.}~\bibnamefont {Ernzerhof}},\ }\href
  {\doibase 10.1103/PhysRevLett.77.3865} {\bibfield  {journal} {\bibinfo
  {journal} {Phys. Rev. Lett.}\ }\textbf {\bibinfo {volume} {77}},\ \bibinfo
  {pages} {3865} (\bibinfo {year} {1996}{\natexlab{a}})}\BibitemShut {NoStop}%
\bibitem [{\citenamefont {Bl\"{o}chl}(1994)}]{Blochl1994}%
  \BibitemOpen
  \bibfield  {author} {\bibinfo {author} {\bibfnamefont {P.~E.}\ \bibnamefont
  {Bl\"{o}chl}},\ }\href {\doibase 10.1103/PhysRevB.50.17953} {\bibfield
  {journal} {\bibinfo  {journal} {Phys. Rev. B}\ }\textbf {\bibinfo {volume}
  {50}},\ \bibinfo {pages} {17953} (\bibinfo {year} {1994})}\BibitemShut
  {NoStop}%
\bibitem [{\citenamefont {Kresse}\ and\ \citenamefont
  {Joubert}(1999)}]{Kresse1999}%
  \BibitemOpen
  \bibfield  {author} {\bibinfo {author} {\bibfnamefont {G.}~\bibnamefont
  {Kresse}}\ and\ \bibinfo {author} {\bibfnamefont {D.}~\bibnamefont
  {Joubert}},\ }\href {\doibase 10.1103/PhysRevB.59.1758} {\bibfield  {journal}
  {\bibinfo  {journal} {Phys. Rev. B}\ }\textbf {\bibinfo {volume} {59}},\
  \bibinfo {pages} {1758} (\bibinfo {year} {1999})}\BibitemShut {NoStop}%
\bibitem [{\citenamefont {Monkhorst}\ and\ \citenamefont
  {Pack}(1976)}]{Monkhorst1976}%
  \BibitemOpen
  \bibfield  {author} {\bibinfo {author} {\bibfnamefont {H.~J.}\ \bibnamefont
  {Monkhorst}}\ and\ \bibinfo {author} {\bibfnamefont {J.~D.}\ \bibnamefont
  {Pack}},\ }\href {\doibase 10.1103/PhysRevB.13.5188} {\bibfield  {journal}
  {\bibinfo  {journal} {Phys. Rev. B}\ }\textbf {\bibinfo {volume} {13}},\
  \bibinfo {pages} {5188} (\bibinfo {year} {1976})}\BibitemShut {NoStop}%
\bibitem [{\citenamefont {Taylor}\ and\ \citenamefont
  {Bruneval}(2011)}]{Taylor2011}%
  \BibitemOpen
  \bibfield  {author} {\bibinfo {author} {\bibfnamefont {S.~E.}\ \bibnamefont
  {Taylor}}\ and\ \bibinfo {author} {\bibfnamefont {F.}~\bibnamefont
  {Bruneval}},\ }\href {\doibase 10.1103/PhysRevB.84.075155} {\bibfield
  {journal} {\bibinfo  {journal} {Phys. Rev. B}\ }\textbf {\bibinfo {volume}
  {84}},\ \bibinfo {pages} {075155} (\bibinfo {year} {2011})}\BibitemShut
  {NoStop}%
\bibitem [{\citenamefont {Guyer}\ \emph {et~al.}(2009)\citenamefont {Guyer},
  \citenamefont {Wheeler},\ and\ \citenamefont {Warren}}]{Guyer2009}%
  \BibitemOpen
  \bibfield  {author} {\bibinfo {author} {\bibfnamefont {J.~E.}\ \bibnamefont
  {Guyer}}, \bibinfo {author} {\bibfnamefont {D.}~\bibnamefont {Wheeler}}, \
  and\ \bibinfo {author} {\bibfnamefont {J.~A.}\ \bibnamefont {Warren}},\
  }\href {\doibase 10.1109/MCSE.2009.52} {\bibfield  {journal} {\bibinfo
  {journal} {Comput. Sci. Eng.}\ }\textbf {\bibinfo {volume} {11}},\ \bibinfo
  {pages} {6} (\bibinfo {year} {2009})}\BibitemShut {NoStop}%
\bibitem [{\citenamefont {Hamel}\ \emph {et~al.}(2008)\citenamefont {Hamel},
  \citenamefont {Williamson}, \citenamefont {Wilson}, \citenamefont {Gygi},
  \citenamefont {Galli}, \citenamefont {Ratner},\ and\ \citenamefont
  {Wack}}]{Hamel2008}%
  \BibitemOpen
  \bibfield  {author} {\bibinfo {author} {\bibfnamefont {S.}~\bibnamefont
  {Hamel}}, \bibinfo {author} {\bibfnamefont {A.~J.}\ \bibnamefont
  {Williamson}}, \bibinfo {author} {\bibfnamefont {H.~F.}\ \bibnamefont
  {Wilson}}, \bibinfo {author} {\bibfnamefont {F.}~\bibnamefont {Gygi}},
  \bibinfo {author} {\bibfnamefont {G.}~\bibnamefont {Galli}}, \bibinfo
  {author} {\bibfnamefont {E.}~\bibnamefont {Ratner}}, \ and\ \bibinfo {author}
  {\bibfnamefont {D.}~\bibnamefont {Wack}},\ }\href {\doibase
  10.1063/1.2839332} {\bibfield  {journal} {\bibinfo  {journal} {Appl. Phys.
  Lett.}\ }\textbf {\bibinfo {volume} {92}},\ \bibinfo {pages} {043115}
  (\bibinfo {year} {2008})}\BibitemShut {NoStop}%
\bibitem [{\citenamefont {Pilania}\ and\ \citenamefont
  {Ramprasad}(2012)}]{Pilania2012}%
  \BibitemOpen
  \bibfield  {author} {\bibinfo {author} {\bibfnamefont {G.}~\bibnamefont
  {Pilania}}\ and\ \bibinfo {author} {\bibfnamefont {R.}~\bibnamefont
  {Ramprasad}},\ }\href {\doibase 10.1007/s10853-012-6411-5} {\bibfield
  {journal} {\bibinfo  {journal} {J. Mater. Sci.}\ }\textbf {\bibinfo {volume}
  {47}},\ \bibinfo {pages} {7580} (\bibinfo {year} {2012})}\BibitemShut
  {NoStop}%
\bibitem [{\citenamefont {Nunes}\ and\ \citenamefont
  {Gonze}(2001)}]{Nunes2001}%
  \BibitemOpen
  \bibfield  {author} {\bibinfo {author} {\bibfnamefont {R.}~\bibnamefont
  {Nunes}}\ and\ \bibinfo {author} {\bibfnamefont {X.}~\bibnamefont {Gonze}},\
  }\href {\doibase 10.1103/PhysRevB.63.155107} {\bibfield  {journal} {\bibinfo
  {journal} {Phys. Rev. B}\ }\textbf {\bibinfo {volume} {63}},\ \bibinfo
  {pages} {155107} (\bibinfo {year} {2001})}\BibitemShut {NoStop}%
\bibitem [{\citenamefont {Souza}\ \emph {et~al.}(2002)\citenamefont {Souza},
  \citenamefont {\'{I}\~{n}iguez},\ and\ \citenamefont
  {Vanderbilt}}]{Souza2002}%
  \BibitemOpen
  \bibfield  {author} {\bibinfo {author} {\bibfnamefont {I.}~\bibnamefont
  {Souza}}, \bibinfo {author} {\bibfnamefont {J.}~\bibnamefont
  {\'{I}\~{n}iguez}}, \ and\ \bibinfo {author} {\bibfnamefont {D.}~\bibnamefont
  {Vanderbilt}},\ }\href {\doibase 10.1103/PhysRevLett.89.117602} {\bibfield
  {journal} {\bibinfo  {journal} {Phys. Rev. Lett.}\ }\textbf {\bibinfo
  {volume} {89}},\ \bibinfo {pages} {117602} (\bibinfo {year}
  {2002})}\BibitemShut {NoStop}%
\bibitem [{\citenamefont {Komsa}\ \emph {et~al.}(2012)\citenamefont {Komsa},
  \citenamefont {Rantala},\ and\ \citenamefont {Pasquarello}}]{Komsa2012}%
  \BibitemOpen
  \bibfield  {author} {\bibinfo {author} {\bibfnamefont {H.-P.}\ \bibnamefont
  {Komsa}}, \bibinfo {author} {\bibfnamefont {T.~T.}\ \bibnamefont {Rantala}},
  \ and\ \bibinfo {author} {\bibfnamefont {A.}~\bibnamefont {Pasquarello}},\
  }\href {\doibase 10.1103/PhysRevB.86.045112} {\bibfield  {journal} {\bibinfo
  {journal} {Phys. Rev. B}\ }\textbf {\bibinfo {volume} {86}},\ \bibinfo
  {pages} {045112} (\bibinfo {year} {2012})}\BibitemShut {NoStop}%
\bibitem [{\citenamefont {Resta}(1977)}]{Resta1977}%
  \BibitemOpen
  \bibfield  {author} {\bibinfo {author} {\bibfnamefont {R.}~\bibnamefont
  {Resta}},\ }\href {\doibase 10.1103/PhysRevB.16.2717} {\bibfield  {journal}
  {\bibinfo  {journal} {Phys. Rev. B}\ }\textbf {\bibinfo {volume} {16}},\
  \bibinfo {pages} {2717} (\bibinfo {year} {1977})}\BibitemShut {NoStop}%
\bibitem [{\citenamefont {Kagimura}\ \emph {et~al.}(2007)\citenamefont
  {Kagimura}, \citenamefont {Nunes},\ and\ \citenamefont
  {Chacham}}]{Kagimura2007}%
  \BibitemOpen
  \bibfield  {author} {\bibinfo {author} {\bibfnamefont {R.}~\bibnamefont
  {Kagimura}}, \bibinfo {author} {\bibfnamefont {R.}~\bibnamefont {Nunes}}, \
  and\ \bibinfo {author} {\bibfnamefont {H.}~\bibnamefont {Chacham}},\ }\href
  {\doibase 10.1103/PhysRevLett.98.026801} {\bibfield  {journal} {\bibinfo
  {journal} {Phys. Rev. Lett.}\ }\textbf {\bibinfo {volume} {98}},\ \bibinfo
  {pages} {026801} (\bibinfo {year} {2007})}\BibitemShut {NoStop}%
\bibitem [{\citenamefont {Hong}\ \emph {et~al.}(2010)\citenamefont {Hong},
  \citenamefont {Kim}, \citenamefont {Lee}, \citenamefont {Shin},\ and\
  \citenamefont {Chung}}]{Hong2010}%
  \BibitemOpen
  \bibfield  {author} {\bibinfo {author} {\bibfnamefont {K.-H.}\ \bibnamefont
  {Hong}}, \bibinfo {author} {\bibfnamefont {J.}~\bibnamefont {Kim}}, \bibinfo
  {author} {\bibfnamefont {J.~H.}\ \bibnamefont {Lee}}, \bibinfo {author}
  {\bibfnamefont {J.}~\bibnamefont {Shin}}, \ and\ \bibinfo {author}
  {\bibfnamefont {U.-I.}\ \bibnamefont {Chung}},\ }\href {\doibase
  10.1021/nl904282v} {\bibfield  {journal} {\bibinfo  {journal} {Nano Lett.}\
  }\textbf {\bibinfo {volume} {10}},\ \bibinfo {pages} {1671} (\bibinfo {year}
  {2010})}\BibitemShut {NoStop}%
\bibitem [{\citenamefont {Ragnarsson}\ and\ \citenamefont
  {Lundgren}(2000)}]{Ragnarsson2000}%
  \BibitemOpen
  \bibfield  {author} {\bibinfo {author} {\bibfnamefont {L.-{\AA}.}\
  \bibnamefont {Ragnarsson}}\ and\ \bibinfo {author} {\bibfnamefont
  {P.}~\bibnamefont {Lundgren}},\ }\href {\doibase 10.1063/1.373759} {\bibfield
   {journal} {\bibinfo  {journal} {J. Appl. Phys.}\ }\textbf {\bibinfo {volume}
  {88}},\ \bibinfo {pages} {938} (\bibinfo {year} {2000})}\BibitemShut
  {NoStop}%
\bibitem [{\citenamefont {Perdew}\ \emph
  {et~al.}(1996{\natexlab{b}})\citenamefont {Perdew}, \citenamefont
  {Ernzerhof},\ and\ \citenamefont {Burke}}]{Perdew1996a}%
  \BibitemOpen
  \bibfield  {author} {\bibinfo {author} {\bibfnamefont {J.~P.}\ \bibnamefont
  {Perdew}}, \bibinfo {author} {\bibfnamefont {M.}~\bibnamefont {Ernzerhof}}, \
  and\ \bibinfo {author} {\bibfnamefont {K.}~\bibnamefont {Burke}},\ }\href
  {\doibase 10.1063/1.472933} {\bibfield  {journal} {\bibinfo  {journal} {J.
  Chem. Phys.}\ }\textbf {\bibinfo {volume} {105}},\ \bibinfo {pages} {9982}
  (\bibinfo {year} {1996}{\natexlab{b}})}\BibitemShut {NoStop}%
\bibitem [{\citenamefont {Broqvist}\ \emph {et~al.}(2008)\citenamefont
  {Broqvist}, \citenamefont {Alkauskas},\ and\ \citenamefont
  {Pasquarello}}]{Broqvist2008}%
  \BibitemOpen
  \bibfield  {author} {\bibinfo {author} {\bibfnamefont {P.}~\bibnamefont
  {Broqvist}}, \bibinfo {author} {\bibfnamefont {A.}~\bibnamefont {Alkauskas}},
  \ and\ \bibinfo {author} {\bibfnamefont {A.}~\bibnamefont {Pasquarello}},\
  }\href {\doibase 10.1103/PhysRevB.78.075203} {\bibfield  {journal} {\bibinfo
  {journal} {Phys. Rev. B}\ }\textbf {\bibinfo {volume} {78}},\ \bibinfo
  {pages} {075203} (\bibinfo {year} {2008})}\BibitemShut {NoStop}%
\end{thebibliography}%

\end{document}